\documentclass[12pt]{article}
\usepackage{geometry}  
\usepackage{graphicx}
\usepackage{amssymb}
\usepackage{amsmath}
\usepackage{epstopdf}
\usepackage{comment}

\usepackage[hyperindex=true,
          pdfstartview=FitH,
          bookmarksnumbered=true,
          bookmarksopen=true,
          citecolor=blue,
          linkcolor=blue,
          colorlinks=true,
          unicode]{hyperref}

\allowdisplaybreaks

\begin{document}

\title{Holographic fluid from nonminimally coupled scalar-tensor theory
of gravity}
\author{Bin Wu and Liu Zhao \\
School of Physics, Nankai University, Tianjin 300071, China\\
{\em email}: \href{mailto:binfen.wu@gmail.com}{binfen.wu@gmail.com} and \href{mailto:lzhao@nankai.edu.cn}{lzhao@nankai.edu.cn}}
\date{}
\maketitle

\begin{abstract}
We establish the gravity/fluid correspondence in the nonminimally coupled
scalar-tensor theory
of gravity. Imposing Petrov-like boundary conditions over the
gravitational field, we find that, for a certain class of background metrics,
the boundary fluctuations obey the standard Navier-Stokes equation for an
incompressible fluid without any external force term in the leading order
approximation under the near horizon expansion. That is to say, the
scalar field fluctuations does not contribute in the leading order
approximation regardless of what kind of boundary condition we impose on it.
\end{abstract}

\section{Introduction}
The AdS/CFT correspondence \cite{Ma} is a successful idea which make a
connection between quantum field theory on the boundary and gravity theory in
the bulk. It has been studied extensively for nearly two decades and has led to
important applications in certain condensed matter problems such as
superconductivity \cite{SC}, etc. In the long wavelength limit, the dual
theory on the boundary reduces to a hydrodynamic
system \cite{PPS,Ba2}, and the transport coefficients of
dual relativistic fluid  was calculated in \cite{Haack:2008cp}.
This is known as gravity/fluid correspondence. 

In analogy to the AdS/CFT duality, the dual fluid usually lives on the AdS 
boundary at asymptotically infinity \cite{Eling:2009sj,Ba3,Ashok}.
However, the choice of boundary at asymptotically infinity is 
not absolutely necessary \cite{Bredberg:2010ky,Strominger2}. 
Refs. \cite{Cai2,Ling:2013kua} attempted to place the boundary at finite cutoff
in asymptotically AdS spacetime to get the dual fluid. 
An algorithm was presented in \cite{Compere:2011dx} for systematically
reconstructing the perturbative bulk metric to arbitrary order.
For spatially flat spacetime, this method has been widely generalized,
such as topological gravitational Chern-Simons theory \cite{Cai:2012mg},
Einstein-Maxwell gravity \cite{Niu:2011gu}, Einstein-dilaton gravity \cite{Cai}
and higher curvature gravity \cite{Eling,Zou:2013ix}.
For spatially curved spacetimes, imposing Petrov-like
boundary conditions on timelike cutoff spacetime is a good way to realize 
boundary fluid equations \cite{Strominger,Wu:2013kqa,Cai:2013uye}, 
provided the background 
spacetime in non-rotating. In \cite{Bin}, the present authors investigated
the fluid dual of Einstein gravity with a perfect fluid source
using the Petrov-like boundary condition. 

In most of the previously known example cases, the dual fluid equation 
will contain an external force term provided the bulk theory involves a 
matter source \cite{Bin, Ba,Ling,Bai:2012ci}. 
In this paper, we proceed to study the fluid dual of a
nonminimally coupled scalar-tensor theory of gravity. We find that 
the dual fluid equation arising from near
horizon fluctuations around certain class of static background configuration
in this theory does not contain the external force term, because the
contribution from the scalar fluctuations is of higher order in the near horizon
expansion and hence does not enter the leading order approximation.

\section{Nonminimally coupled scalar-tensor theory}

We begin by introducing the nonminimally coupled
scalar-tensor theory of gravity in $(n+2)$-dimensions.
The action is written as
\begin{align*}
I[g,\phi ]=\int \mathrm{d}^{n+2}x\sqrt{-g}\left[ \frac{1}{2}(R-2\Lambda)
-\frac{1}{2}(\nabla \phi )^{2}-\frac{1}{2}\xi R\phi^2 - V(\phi)\right] \;,
\end{align*}
where $\xi$ is a coupling constant.  When $\xi = \frac{n}{4(n+1)}$, the theory
becomes a conformally coupled scalar-tensor theory of gravity. We will not
choose any specific value for $\xi$ in this paper because the construction works
for any $\xi$.  We set $8\pi G=1$ for convenience.

The equations of motion that follow from the action read
\begin{align}
  &G_{\mu\nu} + g_{\mu\nu}\Lambda = T_{\mu\nu},  \label{eq1}\\
  &\nabla_{\mu}\nabla^{\mu} \phi-\xi R\phi - \frac{d V}{d \phi}=0,  \label{eq2}
\end{align}
where
\begin{align*}
 T_{\mu\nu}&= \nabla_{\mu}\phi\nabla_{\nu}\phi
 - \frac{1}{2}g_{\mu\nu}(\nabla \phi)^2
 + \xi [g_{\mu\nu} \Box -\nabla_{\mu}\nabla_{\nu} + G_{\mu\nu}]\phi^2
 -g_{\mu\nu}V(\phi).
\end{align*}
In what follows, it is better to reformulate eq.(\ref{eq1}) in the form
\begin{align}
 G_{\mu\nu} &= \tilde{T}_{\mu \nu},
 \label{eq1prime}
\end{align}
in which we have introduced
\begin{align}
&\tilde{T}_{\mu \nu} = \frac{\nabla_{\mu}\phi\nabla_{\nu}\phi - \frac{1}{2}
g_{\mu\nu}(\nabla \phi)^2
    + \xi [g_{\mu\nu} \Box -\nabla_{\mu}\nabla_{\nu}]\phi^2
    -g_{\mu\nu}(\Lambda + V(\phi))}
    {(1 - \xi \phi^2)}\;. \label{tp}
\end{align}

To realize fluid dual of the above theory, we will consider fluctuations around
metrics of the form
\begin{align*}
\mathrm{d} s^2 = -f(r)\mathrm{d}t^2 + \frac{\mathrm{d}r^2}{f(r)} + r^2\mathrm{d}
\Omega_k^2\;,
\end{align*}
where $\mathrm{d}\Omega_k^2$ is the line element of an $n$-dimensional
maximally symmetric Einstein space (with coordinates $x^i$), whose normalized constant sectional
curvature is $k=0,\pm1$. Exact solutions of this form are not yet explicitly
known in arbitrary dimensions. However, a number of example cases indicate
that solutions of the above form indeed exist in some concrete dimensions
\cite{MTZ, Wei, Nadalini}, and, in this work, we do not need to make use of the
explicit solution. Thus the spacetime dimension $d=n+2$, the
metric function $f(r)$ and the scalar potential $V(\phi)$ are all kept
unspecified. In Edington-Fenkelstein (EF) coordinates, the metric can be expressed as
\begin{align}
\mathrm{d} s^2 = g_{\mu\nu}\mathrm{d}x^\mu \mathrm{d}x^\nu
=  -f(r) \mathrm{d}u^2 + 2 \mathrm{d}u \mathrm{d}r + r^2\mathrm{d}
\Omega_k^2\;,   \label{metric2}
\end{align}
where $u$ is the light-like EF coordinate. In the following, whenever
$g_{\mu\nu}$ appears, it is meant to be given by (\ref{metric2}) in the 
coordinates $x^\mu=(u,r,x^i)$.

\section{Hypersurface projection and boundary condition}

To construct the fluid dual of the above system, we need to introduce
an appropriate hypersurface and make make projections for some geometric objects
onto the hypersurface. We also need to introduce appropriate boundary condition
on the projection hypersurface. The formulation is basically parallel to the
previous works such as \cite{Strominger, Bin}.

Consider the timelike hypersurface $\Sigma_c$ defined via $r-r_c=0$ with
constant $r_c$. The induced metric $h_{\mu\nu}$ on the hypersurface is related
to the bulk metric $g_{\mu\nu}$ via
\begin{align}
& h_{\mu\nu}=g_{\mu\nu}-n_{\mu}n_{\nu}, \label{indmet}
\end{align}
where $n_{\mu}$ is the unit normal vector of
$\Sigma_c$. For the line element (\ref{metric2})
\begin{align*}
n_{\mu}=(0,\frac{1}{\sqrt{f(r)}},0,\cdots,0), \qquad  n^{\mu}=(\frac{1}{\sqrt{f(r)}}, \sqrt{f(r)}, 0, \cdots, 0).
\end{align*}
It is natural to introduce $x^a =(u, x^i)$ as a intrinsic coordinate system on the
hypersurface. Note that we have adopted two indexing
systems. Greek indices represent bulk tensors, while Latin indices represent
tensors on the hypersurface. In terms of the coordinates $x^a$, it is convenient
to think of the induced metric $h_{\mu\nu}$ on the hypersurface as a metric
tensor $h_{ab}$ defined on the hypersurface --- one just needs to remove the raw 
$h_{\mu r}$ and the column $h_{r\nu}$ -- which are both full of zeros -- from 
$h_{\mu\nu}$. So, in
the following, we will not distinguish $h_{\mu\nu}$ from $h_{ab}$. We will
sometimes encounter objects with mixed indices such as $h_{\mu a}$. Such objects
should of course be understood as components of a bulk tensor.

The line element corresponding to $h_{ab}$ reads
\begin{align}
 \mathrm{d} s_{n+1}^2 &= -f(r_c) \mathrm{d}u^2+r_{c}^{2}\mathrm{d} \Omega_k^2
 \nonumber \\
 &=-(\mathrm{d} x^{0})^2+r_{c}^{2}\mathrm{d} \Omega_k^2
 \nonumber \\
 &=-\frac{1}{\lambda^2}\mathrm{d}\tau^2 +r_{c}^{2}\mathrm{d} \Omega_k^2,
 \label{ndlelm}
\end{align}
where we have introduced two rescaled temporal coordinates $x^0$ and $\tau$,
which are related to $u$ via $\tau = \lambda x^0 = \lambda \sqrt{f(r_c)}\,u$.
The rescaling parameter $\lambda$ is introduced so that when $\lambda\rightarrow
0$, the theory becomes non-relativistic. It will become clear in the next
section that $\lambda\rightarrow 0$ also signifies the near horizon limit.

The hypersurface projections of eq.(\ref{eq1prime}) can be decomposed into
longitudinal and normal projections, respectively. These two classes of
projections are also known as momentum and Hamiltonian constraints in the case
of pure general relativity. The results of the projections reads
\begin{align}
& D_a(K^{a}{ }_{b}-h^{a}{ }_{b}K)=\tilde{T}_{\mu\nu}n^{\mu}h^{\nu}{ }_{b},
\label{mo}\\
& \hat{R}+K^{ab}K_{ab}-K^2 = - 2\tilde{T}_{\mu\nu} n^{\mu}
n^{\nu}, \label{ha}
\end{align}
where $K_{ab}$ is the external curvature.
The boundary condition to be imposed on the hypersurface is the Petrov-like
condition
\begin{align}
&  C_{(l)i(l)j}=l^{\mu}m_{i}{}^{\nu}l^{\sigma}m_{j}{}^{\rho}C_{\mu\nu\sigma\rho}
=0,
\label{bdry}
\end{align}
where $C_{\mu\nu\sigma\rho}$ is the Weyl curvature tensor, $l^\mu, m^\mu$
together with $k^\mu$ form a set of Newman-Penrose basis vector fields which
obey
\begin{align*}
&  l^2=k^2=0,\,(k,l)=1,\,(l,m_{i})=(k,m_{i})=0,\,(m_{i},m_{j})
=\delta^{i}{ }_{j}.
\end{align*}

The boundary degrees of freedom for the gravitational field are totally encoded
in the Brown-York tensor defined in \cite{B-Y},
\begin{align}
t_{ab}=h_{ab}K-K_{ab},  \label{B-Y}
\end{align}
which has $\frac{1}{2}(n+1)(n+2)$ independent components. The Petrov-like
conditions impose $\frac{1}{2}n(n+1)-1$ constraints over such degrees of freedom,
where the $-1$ is because of the tracelessness of the Weyl tensor. So, there
remain only $n+2$ degrees of freedom, which can be interpreted as the density,
pressure and velocity components of the boundary fluid, which must
obey the Hamiltonian and momentum constraints described earlier in Section 2.
These constraint equations can be viewed as the equation of state and the
evolution equation of the boundary fluid.

Inserting the relation (\ref{B-Y}) into (\ref{bdry}) and making use of
(\ref{mo}) and (\ref{ha}), the boundary condition becomes
\begin{align}
 0&=\frac{2}{\lambda^2}t^{\tau}{ }_{i}t^{\tau}{ }_{j}
 +\frac{t^2}{n^2}h_{ij}-\frac{t}
 {n}t^{\tau}{ }_{\tau}h_{ij}+t^{\tau}{ }_{\tau}t_{ij}\nonumber\\
 &\quad+2\lambda\partial_{\tau}
 \left(\frac{t}{n}h_{ij}-t_{ij}\right)-\frac{2}{\lambda}
 D_{{(}i}t^{\tau}{ }_{j{)}}-t_{ik}t^{k}
 { }_{j} - \hat{R}_{ij}  \nonumber\\
 &\quad -\frac{1}{n}(\tilde{T}_{\nu\rho}n^{\nu}n^{\rho}
 +\tilde{T}+\tilde{T}_{00}-2\tilde{T}_{\rho 0}
 n^{\rho})h_{ij}+\tilde{T}_{ij} , \label{bcex}
\end{align}
where $t = t^a{}_a$ is the trace of the Brown-York tensor.
The calculation that leads to (\ref{bcex}) is quite lengthy and is
basically identical to what we have done in \cite{Bin}, so we refer the readers
to our previous work for details. To proceed, we will need the explicit form
of the tensor $\tilde{T}_{\mu\nu}$. Using (\ref{tp}), the last line in
(\ref{bcex}) can be rewritten as
\begin{align}
&  -\frac{1}{n}(\tilde{T}_{\nu\rho}n^{\nu}n^{\rho} + \tilde{T}+
\tilde{T}_{00}-2\tilde{T}_{\rho 0}
n^{\rho})h_{ij}+\tilde{T}_{ij}    \nonumber \\
& = -\frac{1}{n(1-\xi\phi^2)} \left[ (1+\frac{n}{2})f\phi^{\prime 2} - (f+n)\xi
\Box\phi^2 - 2\Lambda + 4\Lambda\xi\phi^2 + nV(\phi)
\right]h_{ij}.  \label{tpcp}
\end{align}

\section{Fluctuations around the background and order estimations}

Having now described the field equation and the boundary condition, we turn to
look at the perturbative fluctuations around the background (\ref{metric2})
and make order estimations for all relevant quantities.

Since the boundary degrees of freedom from bulk gravity
are all encoded in the Brown-York tensor,
it is reasonable to start with calculations of the Brown-York tensor in the
background spacetime and then making perturbative expansion around the
background values. As in the previous works, we take the expansion parameter
to be identical to the scaling parameter $\lambda$ appeared in (\ref{ndlelm}),
so that the perturbative limit $\lambda\to 0$ is simultaneously the
non-relativistic limit. The perturbed Brown-York tensor reads
\begin{align}
t^{a}{ }_{b}=\sum^{\infty}_{n=0} \lambda^{n}(t^{a}{}_{b})^{(n)},
\end{align}
where
\begin{align*}
&(t^{\tau}{}_{\tau})^{(0)}
=\frac{n\sqrt{f}}{r},   \\
&(t^{\tau}{ }_{i})^{(0)}=0, \\
&(t^{i}{}_{j})^{(0)}=\left(\frac{1}{2\sqrt{f}}\partial_{r}f
+\frac{(n-1)\sqrt{f}}{r}\right)\delta^{i}{ }_{j}
\end{align*}
are the background values. Taking the trace, we also have
\begin{align*}
&t^{(0)}=\frac{n}{2\sqrt{f}}\partial_{r}{f}+ \frac{n^2 \sqrt{f}}{r}.
\end{align*}
Meanwhile, we also take a near horizon limit, assuming
that there exists an event horizon at the biggest zero $r=r_h$
of the smooth function $f(r)$ and that the hypersurface
$\Sigma_c$ is very close to the horizon
as $r_c-r_{h}= \alpha^2\lambda^2$, where $\alpha$ is a constant which is
introduced to balance the dimensionality and can be fixed later.
Doing so the function $f(r_c)$ and relevant quantities can be expanded near the
horizon, e.g.
\begin{align*}
f(r_c)&= f^{\prime}(r_{h})(r_c-r_h)+ \frac{1}{2}f^{\prime \prime}(r_h)
(r_c-r_h)^2
+\cdots \sim \mathcal{O}(\lambda^2).
\end{align*}

Naturally, the scalar field on hypersurface also gets a perturbative expansion,
\begin{align*}
\phi = \sum^{\infty}_{n=0} \lambda^n \phi^{(n)}\;,
\end{align*}
where $\phi^{(0)}$ corresponds to the original (unperturbed) background,
and for static backgrounds of the form (\ref{metric2}), $\phi^{(0)}$ must be
independent of the coordinates $(\tau, x^i)$.
We can further make the near horizon expansion
\begin{align*}
\phi^{(0)}(r_c) &= \phi^{(0)}(r_h) + \phi^{(0)\prime}(r_h) (r_c - r_h) + \cdots
 \sim \mathcal{O}(\lambda^0).
\end{align*}
It is reasonable to assume that for all $n>0$, $\phi^{(n)}$ are functions of $
(\tau, x^i)$ only and independent of $r_c$, because otherwise we can simply
expand these $r_c$-dependent functions near the horizon and absorb the higher
order terms by a redefinition of $\phi^{(n)}$.

Since $\phi^{(0)}$ is $(\tau,x_i)$-independent and $\phi^{(n)}$ are
$r_c$-independent, we get
\begin{align*}
 \partial_\tau \phi &=\partial_\tau(\phi^{(0)} + \lambda\phi^{(1)} + \cdots)
 \sim \mathcal{O}(\lambda^1), \\
\partial_i \phi &=\partial_i(\phi^{(0)} + \lambda\phi^{(1)} + \cdots)
 \sim \mathcal{O}(\lambda^1), \\
\partial_r \phi  &=\partial_r (\phi^{(0)} + \lambda\phi^{(1)} + \cdots)
\sim \mathcal{O}(\lambda^0).
\end{align*}

The index $\mu$ in $\partial_\mu\phi$ can be raised using
the inverse of the bulk metric $g_{\mu\nu}$. Since we concentrate only on the
perturbations on the hypersurface, we need to work out the behaviors of the bulk
metric around $\Sigma_c$.
The components of the bulk metric around $\Sigma_c$ in
$(\tau,r,x^i)$ coordinate can be written as
\begin{align*}
g_{\tau\tau}|_{\Sigma_c} = -\frac{1}{\lambda^2},
\qquad g_{\tau r}|_{\Sigma_c}= 2\lambda\sqrt{f(r_c)},
\qquad g_{ij}|_{\Sigma_c}=r_c^{2}\mathrm{d} \Omega_k^2.
\end{align*}
Therefore,
\begin{align*}
g^{r\tau}|_{\Sigma_c}= \lambda\sqrt{f(r_c)}\sim \mathcal{O}(\lambda^2),
\quad g^{rr}|_{\Sigma_c}=f(r_c)\sim \mathcal{O}(\lambda^2),
\quad g^{ij}|_{\Sigma_c}=\frac{1}{g_{ij}|_{\Sigma_c}}\sim \mathcal{O}(\lambda^0),
\end{align*}
and hence
\begin{align*}
& \partial^\tau \phi|_{\Sigma_c}
= g^{\tau\nu}|_{\Sigma_c}\partial_\nu \phi|_{\Sigma_c}
= g^{r\tau}|_{\Sigma_c} \partial_r \phi|_{\Sigma_c}
\sim \mathcal{O}(\lambda^2),
\qquad  \partial^i \phi|_{\Sigma_c}
= g^{ij}|_{\Sigma_c}\partial_j \phi|_{\Sigma_c} \sim \mathcal{O}(\lambda^1),  \\
& \partial^r \phi|_{\Sigma_c} = g^{r\nu}{}_{|_{\Sigma_c}}\partial_\nu \phi
= g^{r\tau}|_{\Sigma_c} \partial_\tau \phi|_{\Sigma_c}
+ g^{rr}|_{\Sigma_c} \partial_r \phi|_{\Sigma_c} \sim \mathcal{O}(\lambda^2).
\end{align*}

With the aid of all above analysis, the Petrov-like
boundary condition (\ref{bcex}) can be expanded in power series in $\lambda$,
and at the lowest nontrivial order $\mathcal{O}(\lambda^0)$, we get
\begin{align}
 \frac{\sqrt{f'_h}}{\alpha}t^{i}{ }_{j}^{(1)}
 = 2t^{\tau}{ }_{k}^{(1)}t^{\tau}{ }_{j}^{(1)}h^{ik(0)}
- 2h^{ik(0)}D_{(j}t^{\tau}{ }_{k)}^{(1)}
 + \frac{\sqrt{f'_h}}{n \alpha} t^{(1)} \delta^{i}{ }_{j}
- \hat{R}^{i}{ }_{j} - C_h \delta^i{}_j,  \label{d}
\end{align}
where $h_{ij}^{(0)}=r_h^2 \mathrm{d}\Omega_k^2$ and $C_h$ is a constant
with value
\begin{align*}
C_h= \frac{\left(-2\Lambda + \frac{n f'_h}{r_h}
 + 4\Lambda\xi\left(\phi^{(0)}_h\right)^2
 + nV(\phi^{(0)}_h) - n\xi  f'_h \phi^{(0)}_h\phi^{(0)'}_h
 \right)}{n\left(1-\xi\left(\phi^{(0)}_h\right)^2\right)},
\end{align*}
wherein $f'_h$ and $\phi^{(0)'}_h$ represent the derivative of
$f(r)$ and $\phi^{(0)}(r)$ evaluated at $r_h$.

Our aim is to reduce the momentum constraint (\ref{mo}) into the hydrodynamics
equation on the hypersurface. For this purpose, we also need to make order
estimations for the right hand side (RHS) of (\ref{mo}). Since $g_{\mu\nu}
n^{\mu}h^{\nu}{}_b =0$, we have
\begin{align}
\tilde{T}_{\mu \nu}n^{\mu}h^{\nu}{}_b &= \frac{1}{1-\xi\phi^2}
(\partial_{\mu} \phi \partial_{\nu} \phi  - \xi \nabla_\mu\nabla_\nu \phi^2 )
 n^{\mu}h^{\nu}{}_b  \nonumber \\
& = \frac{1}{1-\xi\phi^2}((1-2\xi)n^{\mu} \partial_{\mu} \phi \partial_{b} \phi
- 2\xi\phi n^{\mu} \nabla_\mu\nabla_b \phi).  \label{RHS}
\end{align}
We see that many terms in $\tilde{T}_{\mu \nu}$ drops off after the hypersurface
projection. This makes the order estimation a lot easier.

To estimate the order of (\ref{RHS}), let us first look at the $\tau$ component.
Since
\begin{align*}
n^{\mu} \partial_{\mu} \phi \partial_{\tau} \phi
& = n_r \partial^r \phi \partial_{\tau} \phi
= \frac{1}{\sqrt{f}} \partial^{r} \phi \partial_{\tau} \phi
\sim \mathcal{O}(\lambda^2) ,   \\
n^{\mu} \nabla_\mu \nabla_\tau \phi &= n^{\mu} (\partial_\mu\partial_\tau \phi
 - \Gamma^\nu{}_{\mu\tau}\partial_\nu \phi)
 =\lambda \partial_\tau \partial_\tau \lambda \phi^{(1)}
 \sim \mathcal{O}(\lambda^2),
\end{align*}
where the fact that $\Gamma^\nu{}_{\mu\tau}=0$ and that $\phi^{(1)}$ is
$r$-independent have been used, we see that $\tilde{T}_{\mu \nu}n^{\mu}
h^{\nu}{}_\tau$ is of order $\mathcal{O}(\lambda^2)$. Similarly,
since
\begin{align*}
 n^{\mu} \partial_{\mu} \phi \partial_i \phi
 & = n_r \partial^r \phi \partial_i \phi
 = \frac{1}{\sqrt{f}} \partial^{r} \phi \partial_i\phi
 \sim \mathcal{O}(\lambda^{2}), \\
n^{\mu} \nabla_\mu \nabla_i \phi &= n^{\mu} (\partial_\mu \partial_i \phi
 - \Gamma^\nu{}_{\mu i} \partial_\nu \phi)
  = \lambda \partial_\tau \partial_i \lambda \phi^{(1)}
 - n^{r} \Gamma^j{}_{r j} \partial_i \lambda \phi^1  \\
 & = \lambda^2 \partial_\tau \partial_i \phi^{(1)}
 - \frac{\lambda \sqrt{f}}{r} \partial_i \phi^{(1)}
 \sim \mathcal{O}(\lambda^2),
\end{align*}
we find that $\tilde{T}_{\mu \nu}n^{\mu}
h^{\nu}{}_i$ is also of order $\mathcal{O}(\lambda^2)$. Putting together, we
conclude that the RHS of (\ref{mo}) is a quantity of order
$\mathcal{O}(\lambda^2)$ in the near horizon expansion.

\section{Fluid dynamics on hypersurface}
In terms of the Brown-York stress tensor, the momentum constraint (\ref{mo})
can be rewritten as
\begin{align}
D_{a}t^{a}{ }_{b}= -\tilde{T}_{\mu \nu} n^{\mu}h^{\nu}{}_b. \label{divnf}
\end{align}
We have shown in the last section that the RHS of the above equation
is $\mathcal{O}(\lambda^2)$ in the near horizon expansion. It remains to
consider the near horizon expansion of the left hand side (LHS).

To begin with, let us look at the temporal component. We have
\begin{align}
 D_{a}t^{a}{ }_{\tau}&=D_{\tau} t^{\tau}{ }_{\tau}+D_{i} t^{i}{ }_{\tau}
 \nonumber \\
 &= D_{\tau} t^{\tau}{ }_{\tau}
 - \frac{1}{\lambda^2}D_{i}(t^{\tau}{ }_{j}h^{ij}).
 \label{taucomp}
\end{align}
The near horizon expansion of each term behave as
\begin{align*}
&  D_{\tau} t^{\tau}{ }_{\tau} = D_{\tau} t^{\tau}{ }_{\tau}^{(1)}+\cdots
\sim \mathcal{O}(\lambda^{1}),  \nonumber \\
&  \frac{1}{\lambda^2}D_{i}(t^{\tau}{ }_{j}h^{ij})
= \frac{1}{\lambda} h^{ij(0)}D_{i}
t^{\tau}{ }_{j}^{(1)}+\cdots \sim \mathcal{O} (\lambda^{-1}).
\end{align*}
So, the leading order term of (\ref{taucomp}) is
$\frac{1}{\lambda} h^{ij(0)}D_{i}t^{\tau}{ }_{j}^{(1)}$
at $\mathcal{O}(\lambda^{-1})$. Since the RHS of (\ref{divnf}) is
of order $\mathcal{O}(\lambda^{2})$, we get the following identity at
the order $\mathcal{O}(\lambda^{-1})$:
\begin{align}
D_{i}t^{\tau i(1)}=0. \label{divfree}
\end{align}
Next we consider the spacial components of the momentum constraint.
The LHS reads
\begin{align*}
& D_{a}t^{a}{ }_{i}=D_{\tau}t^{\tau}{ }_{i}+D_{j}t^{j}{ }_{i}.
\end{align*}
Inserting (\ref{d}) into the above equation and noticing that the constant $C_h$
has no contribution after taking the derivative,
we get the following result at order $\mathcal{O}(\lambda^1)$,
\begin{align}
& D_{\tau}t^{\tau}{ }_{i}^{(1)} + \frac{\alpha}{\sqrt{f'_h}}
\left(2t^{\tau}{ }_{i}^{(1)}
D^{k}t^{\tau}{}_{k}^{(1)}+2t^{\tau j (1)}D_{j}t^{\tau}{ }_{i}^{(1)}-D^{k}
(D_{i}t^{\tau}{ }_{k}^{(1)}+D_{k}t^{\tau}{ }_{i}^{(1)})
-D_{j}\hat {R}^{j}{ }_{i}\right)
+ \frac{1}{n}D_{k}t^{(1)} \nonumber  \\
& = \partial_{\tau}t^{\tau}{ }_{i}^{(1)}
+\frac{\alpha}{\sqrt{f'_h}}\left(2t^{\tau j(1)}D_{j}t^{\tau}{ }_{i}^{(1)}
-D_{k}D^{k}t^{\tau}{ }_{i}^{(1)}-\hat{R}^{k}{ }_{i}t^{\tau}{ }_{k}^{(1)}\right)
+\frac{1}{n}D_{k}t^{(1)}.  \label{bdry2}
\end{align}
Once again, since the RHS of eq.(\ref{divnf})
is $\mathcal{O}(\lambda^2)$ in the near horizon expansion,
we get the following nontrivial equation in the leading order
$\mathcal{O}(\lambda^1)$:
\begin{align}
\partial_{\tau}t^{\tau}{ }_{i}^{(1)}
+\frac{\alpha}{\sqrt{f'_h}}\left(2t^{\tau j(1)}D_{j}t^{\tau}{ }_{i}^{(1)}
-D_{k}D^{k}t^{\tau}{ }_{i}^{(1)}-\hat{R}^{k}{ }_{i}t^{\tau}{ }_{k}^{(1)}\right)
+\frac{1}{n}D_{k}t^{(1)} = 0. \label{hy}
\end{align}
Though it seems surprising, the scalar field indeed makes no contribution
in the leading order, regardless of what kind of boundary condition we impose on
it. Therefore, in the leading order approximation, one need
not consider the scalar field as an independent degree of freedom.

Now using the so-called holographic dictionary
\begin{align}
t^{\tau}{ }_{i}^{(1)}=\frac{v_{i}}{2},
\quad \frac{t^{(1)}}{n}=\frac{p}{2},  \label{notatio}
\end{align}
where the $v_i, {p}$  are respectively the velocity and pressure of the dual
fluid on hypersurface, eq.(\ref{hy}) becomes the standard Navier-Stokes equation
on the curved hypersurface, i.e.
\begin{align*}
\partial_{\tau}v_{i}+D_{k}p+ 2v^{j}D_{j}v_{i}-D_{k}
D^{k}v_{i} - \hat{R}^{m}{ }_{i}v_{m} = 0,
\end{align*}
where we have taken $\alpha = \sqrt{f'_h}$ as part of our convention.
Meanwhile, eq.(\ref{divfree}) becomes
\begin{align*}
D_i v^i =0,
\end{align*}
which can be easily identified to be the incompressibility condition for the
dual fluid.

\section{Concluding remarks}
Imposing Petrov-like boundary conditions on a near horizon hypersurface
we have been able to establish a fluid dual for the nonminimally coupled 
scalar-tensor theory of gravity. The resulting Navier-Stokes equation does not 
contain an external force term, as apposed to most of the
previously known examples
cases. The absence of external force term is due to the fact that 
the fluctuations of the scalar field does not contribute in the lowest
nontrivial order in the near horizon expansion. 
Let us remind that the only previously known case in which the force term is 
missing from the fluid dual of gravity with matter source is \cite{Ling}.
The works presented in \cite{Ling} and the present paper naturally raise
the following question: why is the force term missing in the fluid dual of
some theories of gravity with matter source? Currently we don't have the answer 
at hand but it is worth to pay further attention to understand.

\providecommand{\href}[2]{#2}\begingroup
\footnotesize\itemsep=0pt
\providecommand{\eprint}[2][]{\href{http://arxiv.org/abs/#2}{arXiv:#2}}

\end{document}